\documentclass[pdflatex,sn-mathphys-num]{sn-jnl}

\usepackage{graphicx}%
\usepackage{multirow}%
\usepackage{amsmath,amssymb,amsfonts}%
\usepackage{amsthm}%
\usepackage{mathrsfs}%
\usepackage[title]{appendix}%
\usepackage{xcolor}%
\usepackage{textcomp}%
\usepackage{manyfoot}%
\usepackage{booktabs}%
\usepackage{algorithm}%
\usepackage{algorithmicx}%
\usepackage{algpseudocode}%
\usepackage{listings}%

\theoremstyle{thmstyleone}%
\theoremstyle{thmstyletwo}%

\theoremstyle{thmstylethree}%
\newcommand{\bfo}{BiFeO$_3$~}

\begin{document}

\title[Article Title]{\bfo nanoparticles at low-temperature using atomistic simulations - surface charge distribution and terminations}


\author[1]{\fnm{Mauro A.} \sur{P. Gon\c{c}alves}}\email{goncalves@fzu.cz}
\author[1]{\fnm{Mon\'ica} \sur{Graf}}
\author[1]{\fnm{Marek} \sur{Pa\'sciak}}
\author[1]{\fnm{Ji\v{r}\'i} \sur{Hlinka}}

\affil[1]{\orgdiv{Department of Dielectrics}, \orgname{FZU--Institute of Physics of the Czech Academy of Sciences}, \orgaddress{\street{Na Slovance 2}, \city{Prague}, \postcode{182 00}, \country{Czech Republic}}}

\abstract{
This paper analyzes how the ferroelectric properties of cubic-like \bfo nanoparticles are affected by different terminations and charge distributions at the surface using ab-initio-based atomistic computational experiments. 
Our findings unveil multiple multidomain configurations and illustrate how the different order parameters evolve towards the surface. Interestingly, for neutral terminations, a non-rhombohedral phase of \bfo with a stripe-like polarization arrangement was stabilized. We evaluate the polarization, oxygen octahedra rotation, and volume variation for all the configurations obtained, taking advantage of the atomic-scale details provided by the methods used in this study.
}


\maketitle

\section{Introduction}\label{sec1}


In recent years, several studies have shown that ferroelectric materials can efficiently clean water pollutants with low energy consumption ~\cite{Amdouni2023,Kalhori2022,Xian2011,Mushtaq2018}. Ferroelectric nanoparticles exhibit a large photocatalytic response because of their switchable polarization, which acts as an internal electric field. This field drives photo-exited electrons and holes in opposite directions, reducing their recombination rate ~\cite{Paillard2016}. 
Moreover, photocatalysis can be combined with piezocatalysis to increase pollutant degradation rates ~\cite{Lu2022,Zhou2022,Amdouni2023,Kalhori2022} if the catalytic material is also a good piezoelectric material, as is the case for ferroelectric materials. Piezoelectric materials generate charges under mechanical stress and deform when subjected to an external electric field. 
Piezocatalysis is a new advanced oxidation process where mechanical vibrations, such as acoustic waves, generate waves of pressure at the surface of the ferroelectric nanoparticles. These pressure waves couple with the local polarization at the nanoparticle surface inducing local polarization. Consequently, electric fields are generated, driving the electrons and holes apart, making them available for redox reactions~\cite{Lu2022,Zhou2022,Zhang2022}.

Special attention has been given to \bfo nanoparticles due to their promising results~\cite{Amdouni2023,Mushtaq2018}. 
\bfo is a well-studied lead-free perovskite as well as a multiferroic material in which ferroelectric, ferroelastic, and antiferromagnetic orders are simultaneously present with potential for applications in spintronics, nonvolatile memory devices, and sensors~\cite{wang_2020} and novel smart devices~\cite{chaudron_2024}.
At room temperature, \bfo is in a rhombohedral phase with a spontaneous polarization between $0.6-1.0~C/m^2$ coexisting with the antiferrodistortive modes involving oxygen octahedra rotation about $13.8^\circ~$ in antiphase~\cite{lebeugle_2007}.
Regarding piezocatalytic response, \bfo presents a large polarization of $1.0~C/m^2$, as mentioned before, as well as a high piezoelectric coefficient of $100 pm/V$~\cite{Lebeugle2007}, combined with a dielectric constant of approximately $50$. The combination of these properties provides \bfo with high photocatalytic activity, as recently demonstrated in studies on nanoparticles, nanowires, and nanosheets~\cite{Mushtaq2020}. 
Furthermore, \bfo has a bandgap in the range $2.2 -2.6 eV$~\cite{Lebeugle2007}, which is a relatively small value when compared to other ferroelectric materials. This makes \bfo suitable for applications utilizing sunlight.

In recent work, Wafa Amdouni and collaborators~\cite{Amdouni2023} investigated the piezo-photocatalytic properties of \bfo nanoparticles with $60$ nm size. The results show that in approximately 30 minutes, the \bfo nanoparticles successfully degraded $97\%$ of the Rhodamine B, a model aqueous organic dye pollutant, using only the piezocatalytic effect. 
This result is the highest rate ever reported and 3 times higher than the second-highest rate obtained achieved $MoS_2$ nanosheets~\cite{Li2019}. 
When the ultrasonic vibrations are combined with sunlight, the \bfo nanoparticles present the highest piezo-photocatalytic rate ever reported and $3$ times higher than the previous record observed with $Bi_{4}Ti_{3}O_{12}$ nanoplates~\cite{Xie2022}.

However, the arrangement of polarization and oxygen octahedra rotations inside the \bfo nanoparticles remains unclear. In particular, the impact of different charge distributions at the surface on the nanoparticle's internal atomic structure is unknown.  
At the nanoscale, the homogeneous state of polarization becomes unstable due to multiple factors, including size and shape, mechanical stress, electrical boundary conditions, and others~\cite{Wu2015}. 
Consequently, the polarization inside nanostructures breaks into domains where the polarization points in different directions. This leads to the formation of various complex polarization arrangements, ranging from flux-closure domain structures to vortices and even skyrmions or hopfions~\cite{Franco2020,Naumov2004,Stachiotti2011,Lukyanchuk2020}.

We aim to investigate cubic nanoparticles of \bfo with a size of approximately $16$ \AA~ using ab initio-based atomistic shell model potential optimized for \bfo molecular dynamics simulations~\cite{graf_2014,graf_2014_2}. This model has been successfully applied to reproduce the experimental polarization and oxygen octahedra rotation of \bfo and capture the correct sequence of \bfo phase transitions. 

\begin{figure}[h]
\centering
\includegraphics[width=0.7\textwidth]{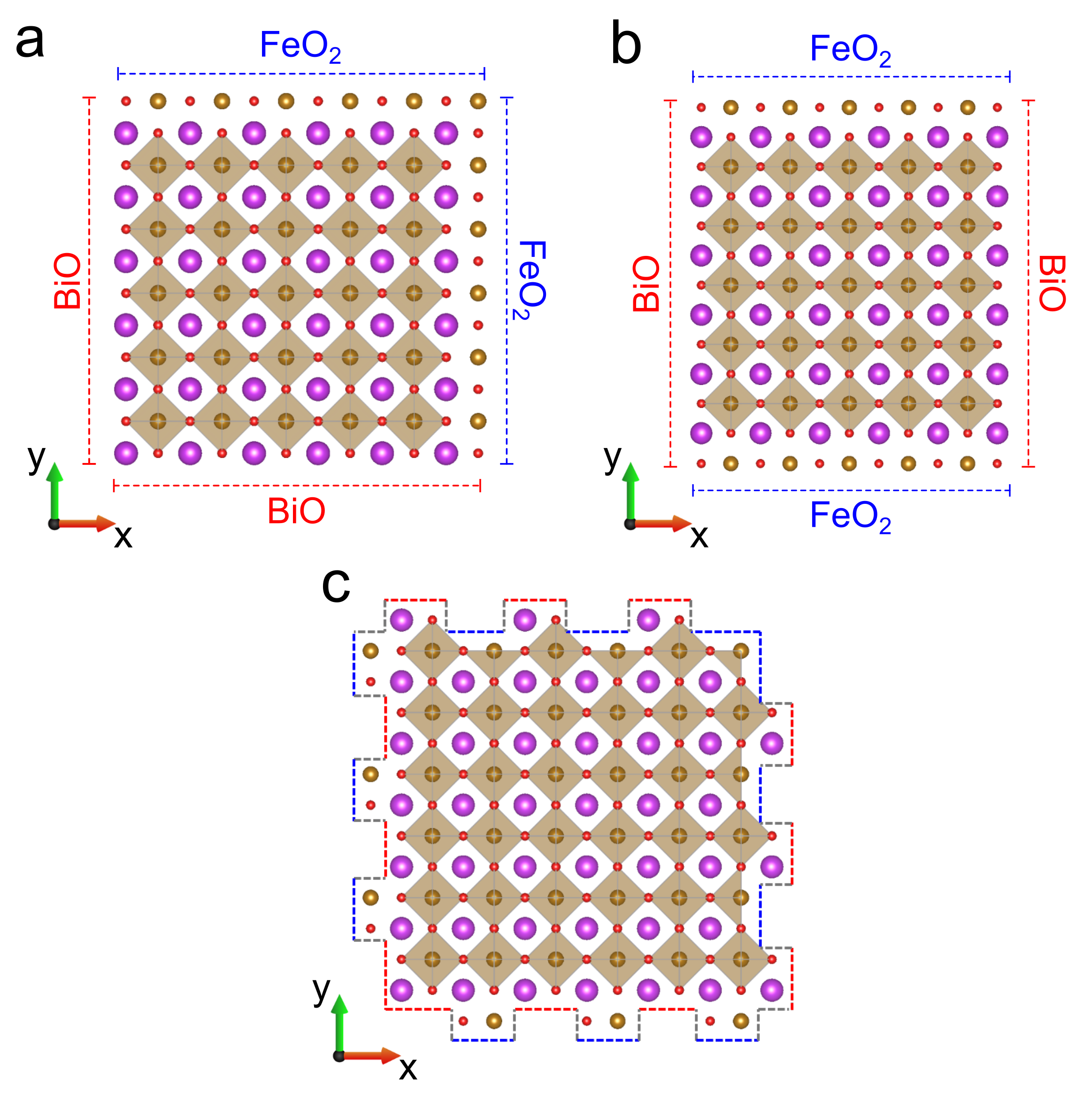}
\caption{Sketch of the different terminations implemented in the construction of the nanoparticle.}
\label{fig_1}
\end{figure}

In the shell model potential, each atom is represented by a core and a shell, each with its respective charge. In the model used in this work, the BiO plane is positively charged, while the FeO$_2$ plane has the exact opposite charge. Therefore, by constructing nanoparticles with different terminations, we simulate various charge distributions at the nanoparticle surface. All the nanoparticles constructed in this work have a null total charge.
The three different sets of nanoparticles investigated in this work are presented in Fig. \ref{fig_1}~ and represent different charge distributions at the nanoparticle surface. For the first nanoparticle, the opposite faces of the cubic-like nanoparticle have opposite charges, as shown in Fig. \ref{fig_1}a. 

The second nanoparticle was built to have opposite faces with the same charges, which means that both faces of the cubic nanoparticle normal to x are terminated with the BiO plane and the faces perpendicular to y with the FeO$_2$ planes (see Fig. \ref{fig_1}b). 
Finally, for the last nanoparticle, we implemented a mixed interface that combines at the same face of the nanoparticle regions with BiO and FeO$_2$ termination, simulating a nanoparticle with a neutral surface, as shown in Fig. \ref{fig_1}c.

\section{Calculation details}\label{cal_details}
To study \bfo nanoparticles, we used an interatomic potential with parameters fitted from ab-initio calculations~\cite{graf_2014,graf_2014_2}. The potential utilizes the framework of the shell model, representing each atom in the system with a core and a shell, mimicking the atomic polarizability and with the anharmonic core-shell interaction is particularly well-suited in reproducing ferroelectric behavior~\cite{Tinte2004}. 
The low-temperature multidomain configuration was stabilized using nanoparticles made of 40$\times$40$\times$40 \bfo cubic cells (\textit{i.e.} 320.000 atoms). In the initial configurations, the domains were set by displacing the Bi atoms by 0.2 \AA~ from their paraelectric ideal position along one of the eight equivalent $\langle 111 \rangle$ directions. 

The structural relaxation was achieved using classical molecular dynamics simulations under constant stress and temperature conditions. A temperature of 1~K was chosen to facilitate effective relaxation while introducing a minimal amount of thermal noise. The simulation began with a 10~ps thermalization, followed by an additional 10~ps during which the average configuration was computed.  
The molecular dynamics calculations were performed using the DL\_POLY software~\cite{Todorov2006}.

Local dipole moments were evaluated for each 5-atom perovskite cell centered in the Bi and Fe atoms based on the positions and charges of cores and shells. The rotations of the oxygen octahedra around the Fe atoms were computed with respect to the center of mass of the oxygen octahedra.
The local volume of the unit cell centered around the Bi was computed from the position of the eight Fe atoms around, and vice versa to compute for the unit cell centered in the Fe atoms.

\section{Results and Discussion}

\subsection{Nanoparticle with opposite faces terminated either with BiO or FeO$_2$}\label{1st_termination}

In the initial cubic-like \bfo nanoparticles, the opposite faces of the nanoparticle are terminated with either BiO or FeO$_2$, as shown in Fig. \ref{fig_1}a. This set of terminations creates distinct positive and negative charge distributions on the faces of the nanoparticle. The BiO-terminated faces exhibit a positive charge, while the FeO$_2$-terminated faces a negative charge. 

\begin{figure}[h]
\centering
\includegraphics[width=0.7\textwidth]{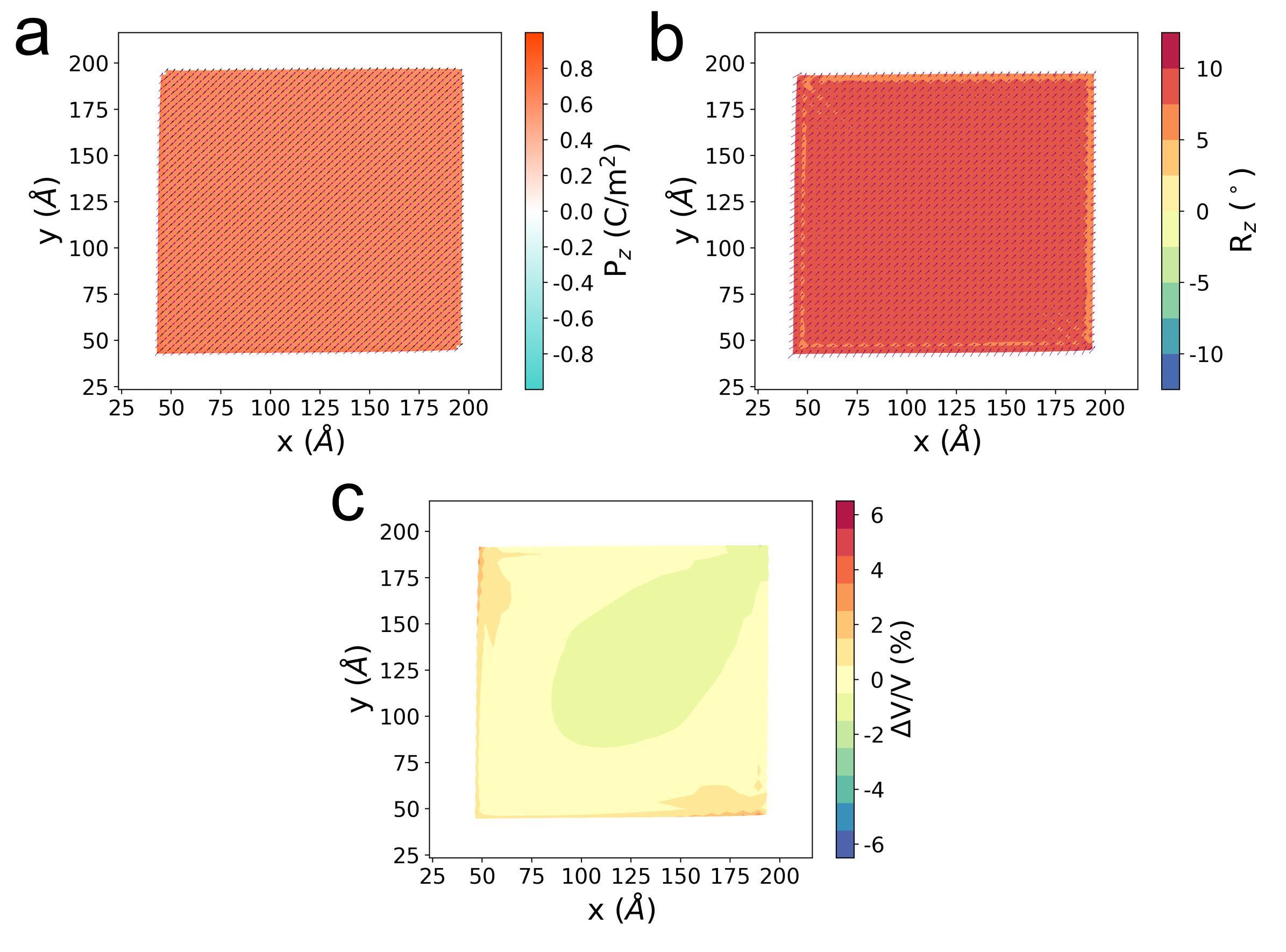}
\caption{Properties of the nanoparticle with opposite faces terminated either with BiO or FeO$_2$. Panel a with the polarization map of multidomain configuration. Panel b with the oxygen octahedra rotation pattern in antiphase. Panel c with the volume deviations with respect to the average unit cell volume. }
\label{fig_2}
\end{figure}

In Fig. \ref{fig_2}a, we show the polarization map obtained for a cut normal to the z-axis through the center of the relaxed nanoparticle. The polarization is homogeneous and parallel to the $\langle 111 \rangle$ direction, consistently pointing from the surface terminated with BiO to the FeO$_2$ surface. The polarization magnitude inside the nanoparticle is constant ($\approx 1~C/m^2$), with small deviations at the surfaces of the nanoparticle.
Similar behavior is observed for the rotations of the oxygen octahedra, presented in Fig. \ref{fig_2}b. Inside the nanoparticle, the rotation is $\approx 13.4^\circ$ along the $\langle 111 \rangle$ direction in anti-phase. However, similar to the polarization arrangement, deviations from this value are observed as we approach the surface. At the BiO termination, the oxygen octahedra closest to the surface exhibit rotations up to $20^\circ$, decreasing to approximately $11.0^\circ$ in the second octahedra before converging to the value of the rotation obtained inside the nanoparticle. At the FeO$_2$ surface, the deviations of the rotations are smaller, exhibiting a slight decrease in magnitude. 

Moreover, we investigated the local atomic structure distortions by computing the relative volume for each perovskite unit cell centered on the Bi and Fe atoms, as shown in Fig. \ref{fig_2}c. At the center of this cut of the nanoparticle, the volume of the unit cell is approximately $1\%$ smaller than the average volume in the nanoparticle. The compressed region extends predominantly along the diagonal of the cubic-like nanoparticle, parallel to the $\langle 111 \rangle$ direction, spanning from the nanoparticle's corner to corner and passing through the center. For the rest of the nanoparticle's surface, the volume increases in some regions up to $6\%$, especially at the vertices and edges of the nanoparticle.

\subsection{Nanoparticle with BiO termination on faces normal to $x$ and FeO$_2$ termination on faces normal $y$}\label{2nd_termination}

The second nanoparticle was constructed with either BiO or FeO$_2$ termination, like in the previous section. However, for this nanoparticle, both faces normal to $x$ are positively charged (BiO termination), while the faces normal to $y$ are negatively charged (FeO$_2$ termination), as presented in Fig. \ref{fig_1}b. The faces perpendicular to $z$ retain the same charge distribution as implemented in the previous section, exhibiting opposite charge distributions on the top and bottom surface of the nanoparticle. 

\begin{figure}[h]
\centering
\includegraphics[width=0.8\textwidth]{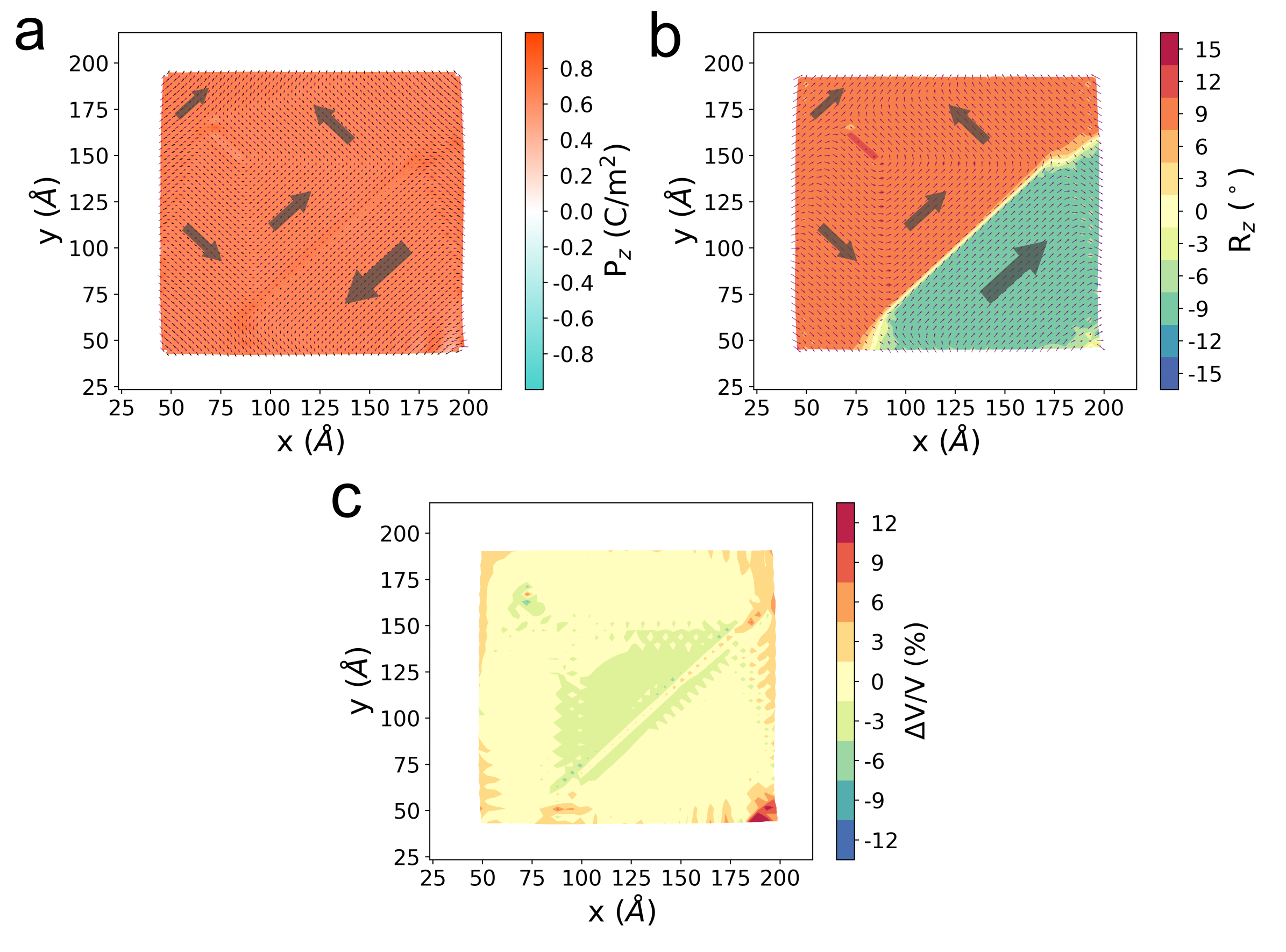}
\caption{Properties of the BiO termination on faces normal to $x$ and FeO$_2$ termination on faces normal $y$. Panel a with the polarization map of multidomain configuration. Panel b with the oxygen octahedra rotation pattern in antiphase. Panel c with the volume deviations with respect to the average unit cell volume. }
\label{fig_3}
\end{figure}

The rearrangement of charges at the surface leads to a completely different arrangement of polarization inside the nanoparticle. 
Fig. \ref{fig_3}a shows the polarization map obtained for a cut through the nanoparticle center, normal to the $z$.
The polarizing along $z$ is always positive, pointing from the bottom (BiO termination) to the top (FeO$_2$ termination). This result agrees with the findings in the previous Section \ref{1st_termination}. 
On the other hand, the polarization in the $xy$ plane breaks into different domains. At the BiO terminations, the polarization points toward the inside of the nanoparticle, while at the FeO$_2$ termination, the polarization points in the opposite direction. 
To accommodate the polarization imposed by the terminations, the polarization breaks into different domains separated by $109^\circ$ and $71^\circ$ domain walls. 

Fig. \ref{fig_3}b shows the map of oxygen octahedra rotation in antiphase, identifying the domain arrangement observed in the polarization map. To minimize the system energy, the octahedra rotations and local polarization should be colinear. Fig. \ref{fig_3}b shows the formation of domains where the oxygen octahedra rotation is either parallel or antiparallel to the local polarization. At the surface of the nanoparticle, the rotations increase in magnitude at the BiO termination while decreasing at the FeO$_2$ termination. 

The analysis of the local volume, presented in Fig. \ref{fig_3}c, shows that the greatest variations in volume are observed on the surface of the nanoparticle, as expected, and around the domain walls. It is important to note that the range of volume variation in Fig. \ref{fig_3}c is twice as large as that presented in Fig. \ref{fig_2}c.

\subsection{Nanoparticle with mixed termiations}\label{3rd_termination}

So far all the nanoparticles investigated were constructed with the same termination in each face. As a consequence, at each face, there is a positive or negative distribution of charge. 
In this section, we will study a nanoparticle with faces with a total charge of zero, constructed by intercalating BiO and FeO$_2$ termination within the same face in a zigzag pattern. 

\begin{figure}[h]
\centering
\includegraphics[width=0.8\textwidth]{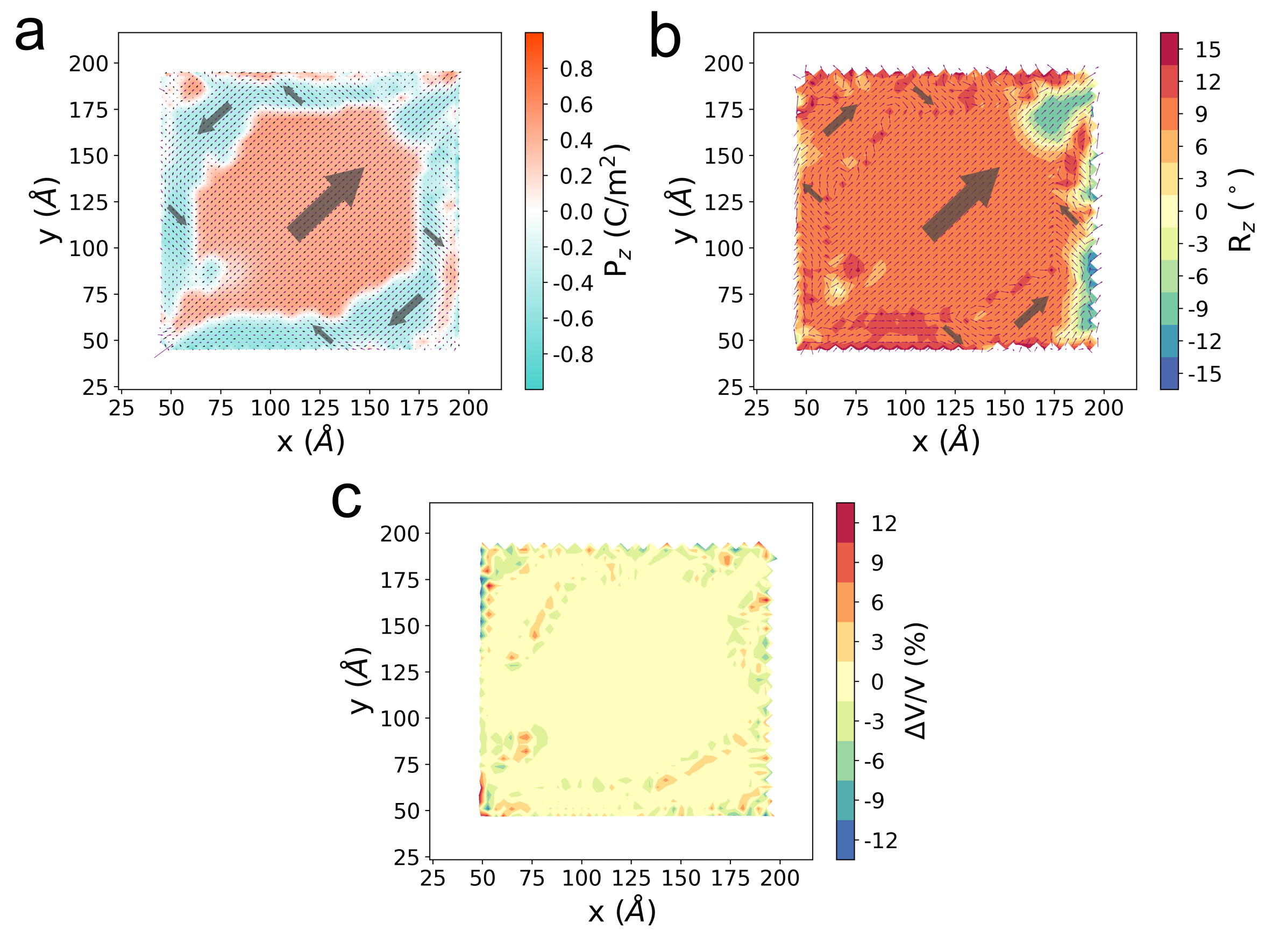}
\caption{Properties of the nanoparticle with mixed termination. Panel a with the polarization map of multidomain configuration. Panel b with the oxygen octahedra rotation pattern in antiphase. Panel c with the volume deviations with respect to the average unit cell volume. }
\label{fig_4}
\end{figure}

The relaxed configuration of the nanoparticle with mixed terminations is presented in Fig. \ref{fig_4}. The polarization inside the nanoparticle breaks into domains, leading to a nanoparticle total polarization of 0 C/m2.
The polarization map, presented in Fig. \ref{fig_4}a, shows at the center of the nanoparticle a large domain with polarization along $\langle 111 \rangle$. Around this domain, the polarization is negative along $z$ and forms in the xy plane to vortices. This pattern can be observed through all the nanoparticles, with the axis of the vortices parallel to the diagonal of the cubic nanoparticle. 
Near the surface of the nanoparticle, the local polarization is strongly affected by the polarization in the atoms around.

The rotation of the oxygen octahedra and the polarization are colinear inside the nanoparticle as shown in Fig. \ref{fig_4}b. However, at the surface of the nanoparticle, the oxygen octahedra rotations increase, in some areas to more than twice the values obtained inside the nanoparticle. 

Finally, analyzing the variations of volume in the nanoparticle, presented in Fig. \ref{fig_4}c, it is possible to identify large deviations from the average unit cell volume at the surface of the unit cell. Away from the surface, inside the nanoparticle, the volume slightly deviates at the domain walls. 

\begin{figure}[h]
\centering
\includegraphics[width=0.8\textwidth]{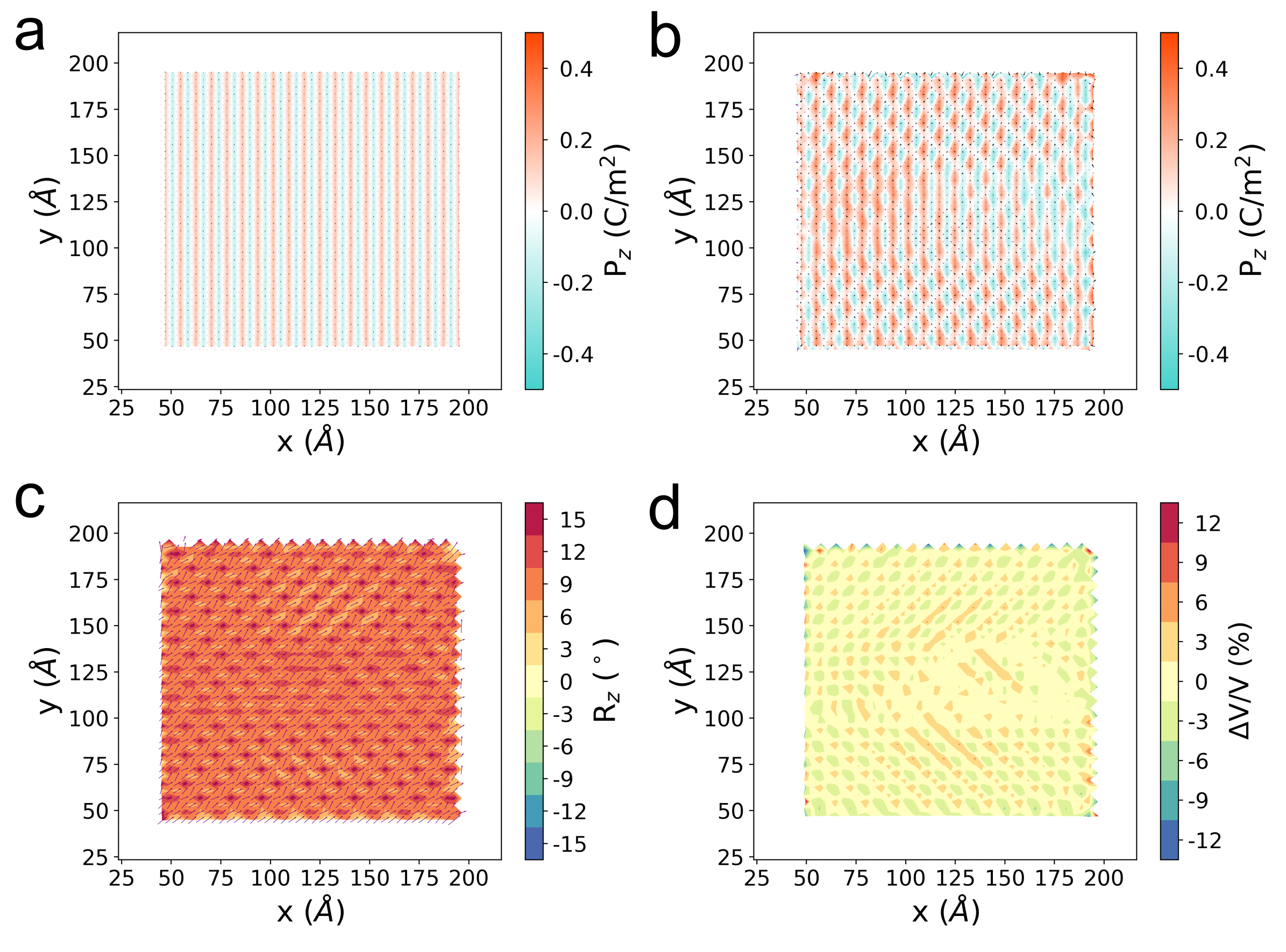}
\caption{Properties of the nanoparticle with mixed termination with the stripe multidomain arrangement . Panel a with the polarization map of multidomain configuration. Panel b with the oxygen octahedra rotation pattern in antiphase. Panel c with the volume deviations with respect to the average unit cell volume.}
\label{fig_5}
\end{figure}

When we analyzed the polarization arrangement in the nanoparticle, in some regions strip-like domains emerged. To stabilize this arrangement of the polarization we construct a new nanoparticle, with the same termination at the surface, but displacing the Bi atoms along the $\langle 111 \rangle$ and $\langle \bar{1}\bar{1}\bar{1} \rangle$ directions forming a stripe-like structure, as shown in Fig. \ref{fig_5}a. 

Fig. \ref{fig_5}b shows the relaxed arrangement of the polarization, with the stripe-like domains modulated along $x$, as prescribed in the initial configuration. The polarization map obtained reveals the development of modulations along $y$ and $z$ on top of the stripes.
Finally, it is important to note that the magnitude of the polarization in the strip-like domains is smaller than that obtained in all the other configurations presented previously. 

In the stripe-like domains, the oxygen octahedra pattern is drastically different from the bulk \bfo. In this configuration, the polarization rotates in phase along the $x$ direction. In Fig 5c the rotations map for the in-phase pattern is presented. The rotation is mainly along the $\langle 111 \rangle$ direction, with small deviations induced by the polarization pattern. 

The local variation of the volume, shown in Fig. \ref{fig_5}d, reflects the local change of polarization in the stripe-like domains, with the oscillation of the volume around the average unit cell volume obtained for the nanoparticle.

\section{Conclusion}\label{Conclusion}

In summary, we have explored how different sets of terminations affect the arrangement of polarization and oxygen octahedra rotations inside \bfo nanoparticles. We observed various polarization arrangements, ranging from a homogeneous monodomain to complex configurations with vortex formations. Interestingly, for nanoparticles with neutral terminations, a stripe-like configuration was stabilized.

\backmatter

\bmhead{Acknowledgments}

This work was supported by the Czech Science Foundation (project no. 19-28594X) and by the Eu- ropean Union?s Horizon 2020 research and innova- tion programme under grant agreement no. 964931 (TSAR). MAPG was supported by the European Union and the Czech Ministry of Education, Youth and Sports (Project: MSCA Fellowship CZ FZU I - CZ.02.01.01/00/22 010/0002906).

\bibliography{bibliography.bib}

\end{document}